\documentclass[prb,twocolumn,groupedaddress,showpacs]{revtex4}
\usepackage{graphicx}
\begin{document}
\bibliographystyle{prbrev}

\title{Desorption of alkali atoms from $^4$He nanodroplets}

\author{Alberto Hernando}
\altaffiliation [Present address:]{
Laboratoire Collisions, Agr\'egats, R\'eactivit\'e, IRSAMC, Universit\'e
Paul Sabatier 118 route de Narbonne 31062 - Toulouse CEDEX 09, France
}
\affiliation{Departament ECM, Facultat de F\'{\i}sica,
and IN$^2$UB,
Universitat de Barcelona. Diagonal 647,
08028 Barcelona, Spain}

\author{Manuel Barranco}
\affiliation{Departament ECM, Facultat de F\'{\i}sica,
and IN$^2$UB,
Universitat de Barcelona. Diagonal 647,
08028 Barcelona, Spain}

\author{Mart\'{\i} Pi}
\affiliation{Departament ECM, Facultat de F\'{\i}sica,
and IN$^2$UB,
Universitat de Barcelona. Diagonal 647,
08028 Barcelona, Spain}

\author{Evgeniy Loginov}
\affiliation{
Laboratoire de Chimie Physique Mol\'eculaire, Swiss Federal Institute of
Technology Lausanne (EPFL), CH-1015 Lausanne, Switzerland
}

\author{Marina Langlet}
\affiliation{
Laboratoire de Chimie Physique Mol\'eculaire, Swiss Federal Institute of
Technology Lausanne (EPFL), CH-1015 Lausanne, Switzerland
}

\author{Marcel Drabbels}
\affiliation{
Laboratoire de Chimie Physique Mol\'eculaire, Swiss Federal Institute of
Technology Lausanne (EPFL), CH-1015 Lausanne, Switzerland
}

\begin{abstract}
The dynamics following the photoexcitation of  Na and Li  atoms
located on the surface of helium nanodroplets has been
investigated in a joint experimental and theoretical study.
Photoelectron spectroscopy has revealed that excitation of the
alkali atoms via the $(n+1)s \leftarrow ns$ transition leads to the desorption
of these atoms. The  mean kinetic energy of the desorbed atoms, as
determined by ion imaging, shows a linear dependence on excitation
frequency. These experimental findings are analyzed within a
three-dimensional, time-dependent density functional approach for
the helium droplet combined with a Bohmian dynamics description of
the desorbing atom. This hybrid method reproduces well the key
experimental observables. The dependence of the observables on
the impurity mass is discussed by comparing the results obtained
for the $^6$Li and $^7$Li isotopes. The calculations show that the
desorption of the excited alkali atom is accompanied by the creation of
highly non-linear density waves in the helium droplet that
propagate at supersonic velocities.


\pacs{36.40.-c, 32.30.Jc, 78.40.-q, 67.40.Yv}

\end{abstract}

\date{\today}

\maketitle

\section{Introduction}

During the last decade, many properties of helium droplets have
been disclosed, especially in relation to their use as a gentle
matrix for spectroscopic experiments. In particular,  the
dynamical aspects associated with the vibrational and rotational
degrees of freedom  of  embedded chromophores, as probed by
infrared spectroscopy, are nowadays well established. We refer the
interested reader to a series of review papers devoted to this
subject.\cite{Kwo00,Sti01,Toe04,Bar06,Sti06,Cho06,Tig07,Sza08} At
variance, the effect of the strong perturbations induced by
electronic excitation or ionization of impurities in helium
droplets is much less understood. For example, there is still no
consistent model that can fully account for the lineshapes and
splittings observed in the electronic spectra of aromatic
molecules.\cite{Bir08} Even less is known about the ensuing
dynamics, like the rearrangement of the helium or the evolution of
photoelectrons.\cite{Log05} These aspects are however expected to
be relevant for the study of chemical reactions in this soft
ultra-cold environment, since chemical reactions involve
essentially the rearrangement of the electronic structure of
reactants. In this respect, understanding the dynamical evolution
of an electronically excited impurity in helium droplets might be
considered a first step towards a better understanding of chemical
reactions in this unique quantum environment.

The aim of the present work is to gain insight into the dynamics
initiated by the excitation of alkali atoms residing on the
surface of helium droplets. The $np \leftarrow ns$ transitions of
alkali-doped helium nanodroplets have been the subject of a series
of experimental and theoretical studies, see {\it e.g.} Refs.
\onlinecite{Sti96,Reh97,Bue07,Her10,Nak01} and references therein.
Only recently, studies involving transitions to higher excited
states have been reported.\cite{Log11a,Log11,Pif10} These studies
reveal that excitation of the alkali atoms in almost all cases
 leads to the desorption of the excited atoms from
the surface of the helium droplet on a picosecond timescale.
\cite{Reh00,Sch01} In the present work we investigate, both
experimentally and theoretically, the desorption process of Na and
Li atoms following excitation to the $4s$ and $3s$ state,
respectively. The use of two different atomic systems allows us to
identify general processes, while the use of lithium having two
light isotopes, $^6$Li and $^7$Li, allows us to quantify the mass
effect on these processes. In the experiments we have used ion and
electron imaging techniques to determine the state distributions
of the desorbed atoms and the velocity distributions of these atoms as function
of excitation frequency. The functional correlation between
excitation energy and final kinetic energy of the impurities has
also been determined from first principles using Fermi's Golden
Rule. The dynamics of the desorption process has been
modeled using a newly developed theoretical hybrid approach based
on time-dependent DFT for the helium density and quantum
trajectories for the impurity wave function.

The hydrodynamic formulation of quantum mechanics --first
developed by Bohm--\cite{bohm} has drawn the attention of
theoretical quantum chemists and physicists because it constitutes
an alternative representation to the time-dependent Schr\"odinger
equation that allows to overcome some of the computational
problems inherent to the conventional quantum mechanical
approach.\cite{qt} The Bohmian formulation bears a large
flexibility, and a variety of quantum trajectory methods are found
in the literature, each of them adapted to the nature of the
dynamical problem under investigation.\cite{qt,qt2,qt3,qt4,Gui95}
We will show how the Bohmian method also lends itself to the study of
the complex dynamical processes addressed in the present work.

This
paper is organized as follows. In Sec. II we describe the
experimental setup. The quantum-trajectory model developed here to
analyze the experimental results is presented in Sec. III.  In
Sec. IV  the experimental results are presented and compared with
the outcome of the calculations. Further remarks on the results
are made in Sec. V, and we conclude with a summary, Sec VI.
Finally, a brief theoretical discussion on the angular dependence
of the velocity distributions is presented in Appendix A, and the
expressions used to simulate the experimental observables are
collected in Appendix B.

\section{Experimental}

The experimental setup has been described in detail
before.\cite{Bra07,Log07} In brief, helium droplets consisting on
average of several thousands of atoms are formed by expanding
high-purity helium gas at a pressure of 30 bar into vacuum through
a 5 $\mu$m orifice cooled to cryogenic temperatures.
 The size distribution of these droplets can be
systematically varied by changing the source
temperature.\cite{Lew93} The helium droplets pick up alkali atoms
as they traverse an oven in which either sodium or lithium  metal
is evaporated. The temperature of the oven is adjusted to ensure
that the droplets on average pick up less than one alkali atom. Via a differential pumping
stage the doped droplets enter a velocity map imaging spectrometer.
 At the center of this setup, the alkali-doped
droplets are excited by crossing the droplet beam perpendicularly
with the frequency-doubled output of a Nd:YAG pumped dye laser
(PrecisionScan SL,  Sirah Laser- und Plasmatechnik GmbH). The
laser system is operated at a repetition frequency of 20 Hz and
provides radiation with a linewidth of less than 0.1 cm$^{-1}$, an
energy of 5 mJ/pulse and a pulse duration of 11$\pm$1 ns. The
laser beam is slightly focused to yield an estimated spot size of
0.37 mm$^2$  at the excitation region. Following excitation, the
alkali atoms are ionized by the absorption of an
additional photon from the same laser pulse. The ions, or
alternatively the photoelectrons, are accelerated by the applied
electric fields and projected onto a position sensitive detector
consisting of a pair of microchannel plates and a phosphor screen.
The light emitted by the phosphor screen is imaged onto a
high-resolution CCD camera (A202k, Basler) that is read out every
laser shot. The individual images are analyzed online and the
centroids of the impacts are determined. The kinetic energy
distributions are determined by performing an inverse Abel
transform on the image constructed from the accumulated centroids.
Spectra are recorded using this setup by monitoring the number of
ion impacts on the detector as function of laser frequency. Both
the ion images and the spectra can be recorded at a specific mass
by gating the front of the detector at the arrival
time of the ions of interest.

\section{Theoretical approach}

The theoretical model is described here in detail for the $4s \leftarrow
3s$ excitation of Na impurities on $^4$He droplets, but  the
formalism applies equally well to the corresponding $3s
\leftarrow 2s$ transition in lithium. We would like to point out that, while the
$(n+1)s \leftarrow ns$ transitions are dipole forbidden in the gas
phase, they are allowed when the atoms reside on the surface of
helium droplets due to the reduced symmetry of the system.\cite{Log11}

Our starting point is the Fermi Golden Rule (GR) for optical
transitions derived from perturbation theory.\cite{Wei80} It
yields the transition probability per unit time $w_{i\rightarrow
f}$ from an initial state $|i\rangle$ to a final state $|f\rangle$
due to the interaction with a perturbative electromagnetic field
$H_I(t)=e^{-i\omega t}V_I$ in a time interval $T$:
\begin{equation}
w_{i\rightarrow f}(\omega)=\frac{1}{\hbar^2T}
\left|\int_Tdte^{-i\omega t}\langle f|e^{i\mathcal{H}t/\hbar}\,
V_I\,e^{-i\mathcal{H}t/\hbar}|i\rangle\right|^2  \; ,
\end{equation}
where $\mathcal{H}$ is the unperturbed hamiltonian that describes
the system. For our purpose here, the initial state is the ground
state (gs) of the complex consisting of a superfluid $^4$He
droplet and a Na atom at zero temperature. We label it as
$|i\rangle=|\Psi_{\mathrm{He}}^{gs},\psi_{\mathrm{Na}}^{gs},\varphi_e^{3s}\rangle$,
where $\Psi_{\mathrm{He}}$ is the nuclear many-body wave function
of the helium cluster, $\psi_{\mathrm{Na}}$ the nuclear wave
function of the sodium atom, and $\varphi_e$ the electronic wave
function of the complex. For convenience, we make explicit that
the Na valence electron is in the nominal $3s$ state, as it is
this electron that interacts with the electromagnetic field. We
are interested in those final states $|f\rangle$ that are
accessible by the optical transition to the nominal 4s state and
we hence label them as $|f\rangle
=|\Psi_{\mathrm{He}}^f,\psi_{\mathrm{Na}}^f,\varphi_e^{4s}\rangle$.

To evaluate the integrand of the GR we make the following
approximations:

\begin{enumerate}
\item[i)] the Born-Oppenheimer (BO) approximation\cite{Wei80} is
used to factorize the electronic and nuclear wavefunctions.
\item[ii)] Density functional theory (DFT)\cite{Dal95,Bar06} is
used to describe the droplet-impurity complex, factorizing the Na
and He nuclear wavefunctions, both evolving according to self-consistent
mean-field hamiltonians.\item[iii)] the Franck-Condon (FC)
principle\cite{Wei80} is invoked, so that it can be assumed that the atomic nuclei do not
change their positions or momenta during the electronic
transition.
\end{enumerate}

\subsection{The electronic contribution}

The BO approximation allows one to factorize the electronic
contribution in the GR expression as
$|\langle\varphi_e^{4s}|V_I|\varphi_e^{3s}\rangle|^2$, and to use
effective pair-potentials for the nuclear hamiltonians. Due to the
dipolar nature of the transition, the angular dependence of
$|\langle\varphi_e^{4s}|V_I|\varphi_e^{3s}\rangle|^2$ can be
written as $P(\theta)= \frac{1}{4\pi}[1+\beta P_2(\cos\theta)]$,\cite{ion,zare}
where $P_2(x)$ is the second Legendre polynomial, $\theta$ is the
angle between the direction of the polarization vector of the
laser light and the final velocity of the adatom, and $\beta$ is the so-called
anisotropy parameter.

Since the projection of the orbital angular momentum onto the
symmetry axis of the system --- defined by the sodium atom and the
center-of-mass of the helium droplet --- does not change during
the electronic transition, the value of $\beta$ can be inferred by
considering the symmetry of the valence electron wavefunctions,
$\varphi_e^{3s}$ and $\varphi_e^{4s}$. Under the assumption that
both these nominally spherical wavefunctions exhibit a dipolar
deformation and that this deformation is larger for the final
wavefunction $\varphi_e^{4s}$ as a result of the stronger
interaction with the helium cluster,\cite{Log11} one finds that
the value of the anisotropy parameter is limited to $1.5<\beta<2$,
see Appendix A.

\subsection{The nuclear contribution within DFT}

For the nuclear contribution, we start from the DFT description of
the system in its ground state as described {\it e.g.} in
Refs.~\onlinecite{Bar06} and \onlinecite{Dal95}.
We want to recall here that this is a zero temperature
phenomenological description
of superfluid liquid $^4$He that reproduces its thermodynamic
properties and elementary excitation spectrum.
The BO
factorization of the electronic wavefunction allows one to represent the
interaction between the helium moiety and the impurity by an
effective He-Na interaction which is based on the
$V_X^{3s}(r_{\mathrm{He-Na}})$ pair-potential, $r_{\mathrm{He-Na}}$ being the
interatomic distance.\cite{Pat91} From the minimization of the energy density
functional $E[\Psi_{\mathrm{He}},\psi_{\mathrm{Na}}]$, we obtain
the effective hamiltonians corresponding to both the helium
particle density
$\rho_{\mathrm{He}}(\mathbf{r})=|\Psi_{\mathrm{He}}(\mathbf{r})|^2$,
and the Na wavefunction $\psi_{\mathrm{Na}}(\mathbf{r})$:
\begin{eqnarray}
H_{\mathrm{He}}^{3s}\Psi_{\mathrm{He}}(\mathbf{r})&\equiv&
\left\{-\frac{\hbar^2 \nabla^2}{2m_{\mathrm{He}}}
 +U[\rho_{\mathrm{He}}]+V_{\mathrm{He}}^{3s}
 (\mathbf{r})\right\}\Psi_{\mathrm{He}}(\mathbf{r})
\nonumber
 \\
&=&\mu\Psi_{\mathrm{He}}(\mathbf{r})
\nonumber
 \\
\nonumber
H_{\mathrm{Na}}^{3s}\psi_{\mathrm{Na}}(\mathbf{r})&\equiv&
\left\{-\frac{\hbar^2 \nabla^2}{2m_{\mathrm{Na}}}+V_{\mathrm{Na}}^{3s}
(\mathbf{r})\right\}
\psi_{\mathrm{Na}}(\mathbf{r})
 \\
&=&\epsilon_{gs}\psi_{\mathrm{Na}}(\mathbf{r})
\label{eqH}
\; .
\end{eqnarray}
Here, $U[\rho_{\mathrm{He}}]$ is the helium DFT self-consistent
potential, $\mu$ the helium chemical potential and $\epsilon_{gs}$
the ground state eigenenergy of the sodium atom in the mean field
created by the helium droplet. The mean field interaction
potentials are obtained by convolution
\begin{eqnarray}
\label{convolutions}
V_{\mathrm{He}}^{3s}(\mathbf{r})&=&\int
d\mathbf{r}'|\psi_{\mathrm{Na}}(\mathbf{r}')|^2V_X^{3s}
(|\mathbf{r}'-\mathbf{r}|)
\nonumber
\\
V_{\mathrm{Na}}^{3s}(\mathbf{r})&=&\int
d\mathbf{r}'|\Psi_{\mathrm{He}}(\mathbf{r}')|^2V_X^{3s}
(|\mathbf{r}'-\mathbf{r}|) \; .
\end{eqnarray}
These density-dependent, mean field hamiltonians are employed to
find the ground state of the droplet-impurity complex ---
$\Psi_{\mathrm{He}}^{gs}$ and $\psi_{\mathrm{Na}}^{gs}$--- used in
the evaluation of the GR. Within the FC approximation
the helium density is frozen, which allows us to identify
$\Psi_{\mathrm{He}}^f$ with $\Psi_{\mathrm{He}}^{gs}$ and to write the total
transition probability as
\begin{equation}
w_{i\rightarrow f}(\omega)\propto[1+\beta P_2(\cos\theta)] \times
I(\omega,f) \nonumber
\end{equation}
with
\begin{equation}
I(\omega,f)=\frac{1}{T}\left|\int_Tdte^{-i(\omega+\omega_{gs}) t}
\langle \psi_{\mathrm{Na}}^f|e^{iH_{\mathrm{Na}}^{4s}t/\hbar}
|\psi_{\mathrm{Na}}^{gs}\rangle\right|^2 \; .
\end{equation}
Here, $\hbar\omega_{gs}=\epsilon_{gs}$, and $H_{\mathrm{Na}}^{4s}$
is the effective hamiltonian for the excited impurity, calculated by evaluating
Eq.~(\ref{convolutions}) using the ground state helium density and the
 the excited state pair-potential $V^{4s}(r_{\mathrm{He-Na}})$ of Ref.~\onlinecite{Pas83}.
Introducing the identity $\sum_n|n\rangle\langle n|= 1$,
where $\left\{|n\rangle\right\}$ is a complete set of
$H_{\mathrm{Na}}^{4s}$ eigenstates with eigenenergies
$\hbar\omega_n$, we obtain
\begin{equation}
\label{eqI}
\begin{array}{rl}
\displaystyle I(\omega,f)=\sum_n&
\displaystyle\left|\langle\psi_{\mathrm{Na}}^f|n\rangle\right|^2\\
&\displaystyle\times\left|\langle
n|\psi_{\mathrm{Na}}^{gs}\rangle\right|^2
\delta\left[\omega-\left(\omega_n-\omega_{gs}\right)\right].
\end{array}
\end{equation}
It is worth noting that summing Eq.~(\ref{eqI}) over the final
states $f$ one obtains the Franck-Condon factors for the Na
absorption spectrum \cite{Her08,Her08b,Lax52}
\begin{equation}
I(\omega)= \sum_f I(\omega,f)=\sum_n\left|\langle
n|\psi_{\mathrm{Na}}^{gs}\rangle\right|^2
\delta\left[\omega-\left(\omega_n-\omega_{gs}\right)\right] \; .
\label{eq6}
\end{equation}
As previously indicated, the final states $\psi_{\mathrm{Na}}^f$
are those accessible by the optical transition to the nominal $4s$
state. By using the same $|n\rangle$ basis for the final states we
recover from Eq. (\ref{eqI}) the Franck-Condon factors. However, in
the ion imaging experiments discussed below what are probed
are adatoms after their desorption from the helium cluster,
which are characterized by a linear momentum $\mathbf{k}$.
For this reason, we are interested in those $|f\rangle$ states that,
after a time $t_\infty$ evolve to asymptotically free states with
well-defined momentum $\mathbf{k}$, namely
$|\psi_{\mathrm{Na}}^f\rangle=U_{\rm Na}(0,t_\infty)|\mathbf{k}\rangle$,
where $U_{\rm Na}(0,t_\infty)$ is the quantum time-evolution operator.

 In this evolution we assume that i) no Na-He exciplexes are formed during the desorption
process, and ii) the helium does not induce relaxation of
the excited adatoms. These assumptions, which are justified \emph{a posteriori}
by the experiments discussed below, imply that all the
states accessible in the course of the excitation evolve to free $4s$ states of Na.
This ensures that the free $4s$ states of Na, $|\mathbf{k}\rangle$,
represent a complete basis.
Introducing these states into Eq. (\ref{eqI})
we obtain
\begin{equation}
\label{eqII}
\begin{array}{rl}
\displaystyle I(\omega,\mathbf{k})=\sum_n&
\displaystyle\left|\langle\mathbf{k}|U_{\rm Na}(t_\infty,0)
|n\rangle\right|^2\\
&\displaystyle\times\left|\langle
n|\psi_{\mathrm{Na}}^{gs}\rangle\right|^2
\delta\left[\omega-\left(\omega_n-\omega_{gs}\right)\right].
\end{array}
\end{equation}
Integrating Eq. (\ref{eqII})
over $\omega$ one obtains the Na probability density in momentum
space after the desorption process
\begin{eqnarray}
I(\mathbf{k})&=& \int d\omega I(\omega,\mathbf{k})\nonumber\\
&=&\left|\langle\mathbf{k}
|U_{\mathrm{Na}}(t_\infty,0)|\psi_{\mathrm{Na}}^{gs}\rangle\right|^2=
|\psi(\mathbf{k},t_\infty)|^2 \; ,
\end{eqnarray}
which is a measured observable in our experiments.
It is worth noting that the evolution defined by $U(0,t_\infty)$
describes the processes that follow the optical excitation.
Hence, at this point the FC approximation is no longer applicable and
the helium density is allowed to evolve dynamically.

\subsection{1D exploratory calculations}

To obtain the dynamical evolution, we have to solve the coupled 3D
time-dependent system
\begin{eqnarray}
i\hbar\frac{\partial}{\partial t} \Psi_{\mathrm{He}}(t,\mathbf{r}) &=&
H_{\mathrm{He}}^{4s}(t) \Psi_{\mathrm{He}}(t,\mathbf{r})
\nonumber
\\
i\hbar\frac{\partial}{\partial t} \psi_{\mathrm{Na}}(t,\mathbf{r}) &=&
H_{\mathrm{Na}}^{4s}(t) \psi_{\mathrm{Na}}(t,\mathbf{r})
\label{coupled}
\end{eqnarray}
from $t=0$ to $t_\infty$, using as initial condition the $3s$ ground
state for both the helium and the Na nuclear wave functions.

\begin{figure}[t]
\includegraphics[width=\linewidth,clip=true]{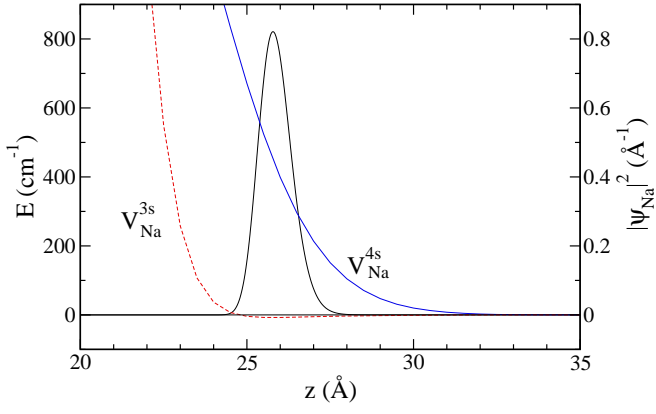}
\caption{\label{fig1} (Color online) Na initial probability
density (thick solid line) used in the 1D calculation and the
$3s$ ground state (dashed line) and $4s$ excited state (solid line)
mean-field potentials. The origin corresponds to the center-of-mass
of the helium droplet.}
\end{figure}

Before attempting the solution of Eqs. (\ref{coupled}), we have
carried out an exploratory 1D evolution. Firstly, we have solved
Eqs. (\ref{eqH}) by means of imaginary time methods --see {\it
e.g.} Refs. \onlinecite{Her08,Her08b} for details-- obtaining the
3D structure of the ground state of a Na@$^4$He$_{1000}$ droplet
(shown in the top left panel of Fig. \ref{fig12}). Secondly,
keeping the helium density frozen, we have let the Na wave
function evolve in the resulting mean field potential along the
symmetry axis, $z$, defined by the center-of-mass of the helium
droplet and the sodium atom. Since the electronic excitation is
assumed to be instantaneous in the FC approximation, the
calculation starts from the Na wave function corresponding to the
ground state of the $V_{\mathrm{Na}}^{3s}$ potential, which next
evolves in the $V_{\mathrm{Na}}^{4s}$ potential shown in
Fig.~\ref{fig1}. We solve the Schr\"odinger equation for Na in a
regular mesh using a predictor-corrector algorithm,\cite{Zwi97}
whose first time steps are provided by the outcome of a
fourth-order Runge-Kutta algorithm. We have found that the Na atom
reaches a quasi-free motion regime after about 1 ps. Since the
helium density is not allowed to evolve, the very large mass of
the helium droplet causes that all the potential energy deposited
into the system during the excitation is converted into kinetic
energy of the Na atom. This is in clear disagreement with the
experimental results discussed below that show that a significant
amount of energy is transferred to the helium droplet. The 1D
calculation is nonetheless useful as it helps to identify
limitations of the approach and it enables us to determine time
and space scales for the full 3D simulations.

\begin{figure}[t]
\includegraphics[width=\linewidth,clip=true]{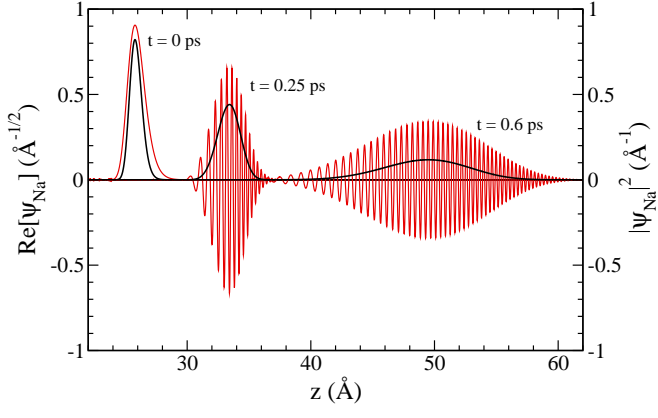}
\caption{\label{fig2} (Color online) Real part of the Na wave
function in the 1D calculation at $t=0$, 0.25, and 0.6 ps. Also
shown are the probability density distributions (thick solid
lines).}
\end{figure}

Due to the highly repulsive nature of the $V_{\mathrm{Na}}^{4s}$
potential, the Na atom leaves the droplet very quickly and attains
a high mean asymptotic velocity of $\sim650$ m/s. As a result of
the high velocity, fast oscillations appear in the Na wave
function as shown in Fig.~\ref{fig2}. In order to reproduce these
high frequency oscillations the use of a very fine spatial mesh is
mandatory. This makes a full 3D evolution computationally
unaffordable. Fortunately, both the probability density and the
velocity field of Na are smooth functions which makes it possible
to describe the full 3D dynamics of the Na atom with quantum
trajectories, an approach that uses positions and velocities
instead of complex wave functions. This allows us to use a hybrid
calculation scheme in which we compute the helium wave function in
a mesh using standard methods for partial differential
equations,\cite{Zwi97} while the evolution of the Na atom is
solved using Bohmian dynamics\cite{qt,bohm} as indicated below. We
mention the existence of other hybrid simulations for the
description of the dynamics of doped helium
droplets.\cite{Wad03,Tak03,Tak04,Bon07,Bon08}

\subsection{Bohmian trajectories for the impurity dynamics}

The equation of motion for the Na trajectories
$\mathbf{R}(t,\mathbf{r})$ is derived as follows.\cite{qt,qt2,qt3,qt4} We
begin with writing the Na wave function in its polar form
$\psi_{\mathrm{Na}}(t,\mathbf{r})= \sqrt{{\cal
R}(t,\mathbf{r})}e^{iS(t,\mathbf{r})/\hbar}$, where ${\cal
R}(t,\mathbf{r})$ is the probability density and $S(t,\mathbf{r})$
is the phase in units of $\hbar$. $S(t,\mathbf{r})$ is also known as the
velocity potential since the velocity field is defined as
$\mathbf{v}(t,\mathbf{r})=\mathbf{\nabla}S(t,\mathbf{r})/m_{\mathrm{Na}}$.
Splitting the Na time-dependent Schr\"odinger equation into its
real and imaginary parts, we obtain a continuity equation for the
probability density coupled to a Hamilton-Jacobi (HJ) equation for
the phase
\begin{equation}\label{split}
\begin{array}{c}
\displaystyle-\frac{\partial {\cal R}(t,\mathbf{r})}{\partial t} =
\mathbf{\nabla}\cdot\mathbf{j}(t,\mathbf{r})
\\
\displaystyle-\frac{\partial S(t,\mathbf{r})}{\partial t} =
\frac{1}{2}m_{\mathrm{Na}}|\mathbf{v}(t,\mathbf{r})|^2 + Q(t,\mathbf{r}) +
V_{\mathrm{Na}}^{4s}(t,\mathbf{r}) \; ,
\end{array}
\end{equation}
where
$\mathbf{j}(t,\mathbf{r})={\cal R}(t,\mathbf{r})\mathbf{v}(t,\mathbf{r})$ is the current
density and
\begin{equation}
Q(t,\mathbf{r})=
-\frac{\hbar^2}{2m_{\mathrm{Na}}}
\frac{\Delta \sqrt{{\cal R}(t,\mathbf{r})}}{\sqrt{{\cal R}(t,\mathbf{r})}}
\end{equation}
is the so-called quantum potential.

Equations (\ref{split}) have been solved as follows. Writing the
density and the current density at time $t$ as an histogram of $M$
test particles with trajectories $\{\mathbf{R}_i(t)\}_{i=1}^M$,
where $\mathbf{R}_i(t)=\mathbf{R}(t,\mathbf{r}_i)$ and
$\mathbf{R}_i(0)=\mathbf{r}_i$, we get
\begin{equation}\label{defRj}
\begin{array}{c}
\displaystyle{\cal R}(t,\mathbf{r})=
\lim_{M\rightarrow\infty}\frac{1}{M}\sum_{i=1}^M\delta[\mathbf{r}-\mathbf{R}_i(t)]
\\
\displaystyle\mathbf{j}(t,\mathbf{r})=
\lim_{M\rightarrow\infty}\frac{1}{M}\sum_{i=1}^M\mathbf{v}[\mathbf{R}_i(t)]
\delta[\mathbf{r}-\mathbf{R}_i(t)]\; .
\end{array}
\end{equation}
The continuity equation is automatically fulfilled if
$\dot{\mathbf{R}}_i(t)=\mathbf{v}[\mathbf{R}_i(t)]$, {\it i.e.,}
if the change in time of the position of the test particle
is just the velocity field evaluated at the position of that test
particle. The equation of motion obeyed by the velocity field
--and thus the equation for the trajectories-- is obtained by
taking the gradient of the HJ equation and rewriting it in the
Lagrangian reference frame ($d/dt=\partial/\partial
t+\mathbf{v}\cdot\mathbf{\nabla}$). One then obtains the quantum Newton
equation
\begin{equation}
m_{\mathrm{Na}}\ddot{\mathbf{R}}_i(t)=-\left.\mathbf{\nabla}\left[Q(t,\mathbf{r})+
V_{\mathrm{Na}}^{4s}(t,\mathbf{r})\right]\right|_{\mathbf{r}=\mathbf{R}_i(t)}
\; .
\end{equation}
In this way, both helium wave function and Na trajectories are
computed consistently at each time step of $10^{-4}$ ps
using a fourth-order Runge-Kutta algorithm. The quantum potential
$Q(t,\mathbf{r})$ is computed using the histogram of the test
particles as probability density
in a regular mesh with  a spatial resolution of $0.35$ \AA,
and 13-points formulas for the
derivatives involved. Simultaneously, the helium wave function is
evolved with a fourth-order Runge-Kutta algorithm, using as
initial condition the $3s$ ground state helium density and a set
of $M=200\,000$ positions randomly generated from the $3s$ ground
state Na probability density. The system is solved up to
$t_\infty=5$ ps ($t_\infty=3$ ps in the case of Li), when the test
particles are far enough from the droplet to follow quasi-free
trajectories at constant velocity. Note that this time is five
times longer than that of the 1D exploratory calculation with the
frozen helium density.

\subsection{Practical evaluation of the Fermi Golden Rule}

To complete the evaluation of the GR expression Eq. (\ref{eqII}),
we have resorted to a semiclassical approximation for the
$H_{\mathrm{Na}}^{4s}$ hamiltonian
for obtaining the dipole absorption
spectrum of atomic impurities:\cite{Her08} the kinetic term is neglected and
the hamiltonian is replaced by the potential energy,
$H_{\mathrm{Na}}^{4s}(0)\rightarrow
V_{\mathrm{Na}}^{4s}(0,\mathbf{r})$. Thus, its eigenstates are
those of the position operator,
$|n\rangle\rightarrow|\mathbf{r}\rangle$, and its eigenvalues
those of the potential energy surface evaluated at  $\mathbf{r}$,
thus $\hbar\omega_n\rightarrow V_{4s}(\mathbf{r})$ and the sum
over states becomes an integral
$\sum_n\rightarrow\int d\mathbf{r}$, allowing us to write
\begin{equation}
\label{eqIb}
\begin{array}{rl}
\displaystyle I(\omega,\mathbf{k})=\int d\mathbf{r}&
\displaystyle\left|\langle\mathbf{k}|U_{\mathrm{Na}}(t_\infty,0)|\mathbf{r}\rangle\right|^2\\
&\displaystyle\times\left|\langle \mathbf{r}|\psi_{\mathrm{Na}}^{gs}\rangle\right|^2
\delta\left[\hbar\omega-\left(V_{4s}(\mathbf{r})-\hbar\omega_{gs}\right)\right].
\end{array}
\end{equation}
Within the quantum trajectory description for the Na atom, the probability
$|\langle\mathbf{k}|U_{\rm Na}(t_\infty,0)|\mathbf{r}\rangle|^2$
is written as
$\delta\left[\mathbf{k}-\mathbf{K}(t_\infty,\mathbf{r})\right]$,
where $\mathbf{K}(t_\infty,\mathbf{r})$ is the momentum at
$t_\infty$ of the trajectory with initial position $\mathbf{r}$ at
$t=0$. Since Na moves at $t_\infty$ as a free particle, we can
safely consider $\hbar\mathbf{K}(t_\infty,\mathbf{r})=
m_{\mathrm{Na}}\dot{\mathbf{R}}(t_\infty,\mathbf{r})$. Using the
definition in Eq. (\ref{defRj}) for the Na probability density,
Eq. (\ref{eqIb}) is then computed as
\begin{equation}
\label{eqIi}
\begin{array}{rl}
\displaystyle I(\omega,\mathbf{k})=\frac{1}{M}\sum_{i=1}^M
&\displaystyle\delta\left[\hbar\mathbf{k}
 -m_{\mathrm{Na}}\dot{\mathbf{R}}_i(t_\infty)\right]\\
&\displaystyle\times\delta\left\{\hbar\omega-
\left[V_{\mathrm{Na}}^{4s}(0,\mathbf{R}_i(0))-\hbar\omega_{gs}\right]\right\}
\; .
\end{array}
\end{equation}
The semiclassical approximation incorporates the trajectories in a
natural way by correlating along each trajectory its initial
potential energy with its asymptotic linear momentum. Note that
though the evaluation of the GR is semiclassical, the trajectories
are quantum-mechanically determined.

Incorporating shape fluctuations of the helium droplet around the
impurity has proved to be crucial for achieving a quantitative
description of processes involving impurities in helium, as they
substantially contribute to the broadening of observables \cite{Her10,Her11,Mat11b}.
For this reason, we have included
the effect of fluctuations in the evaluation of Eq.~(\ref{eqIi})
\emph{a posteriori}, \emph{i.e.}, after the time evolution, using the
DF-sampling method, as shown in the
Appendix B, Eq. (\ref{eqIdis4}). Inclusion of the density fluctuations
mainly affects
the width of physical observables such as the excitation spectrum
and the velocity and kinetic energy distributions of the desorbed atoms.

We want to stress that the transition probability $w_{i\rightarrow
f}$ contains all the physical information that can be experimentally
determined. In point of fact, i) by integrating over $\mathbf{k}$
we obtain the excitation spectrum, as seen in Eq.~(\ref{eq6}); ii)
by fixing the excitation energy $\omega$, we obtain the
distribution over momentum $\mathbf{k}$, corresponding to the
velocity distribution of the desorbed atom; and iii) by the change
of variable $E_k=\hbar^2k^2/2m_{\mathrm{Na}}$ the kinetic energy
distribution is obtained. The explicit expressions used to compute
these quantities are reported in Appendix B.

\section{Results}

\subsection{Excitation spectra}

\begin{figure}[t]
\includegraphics[width=\linewidth,clip=true]{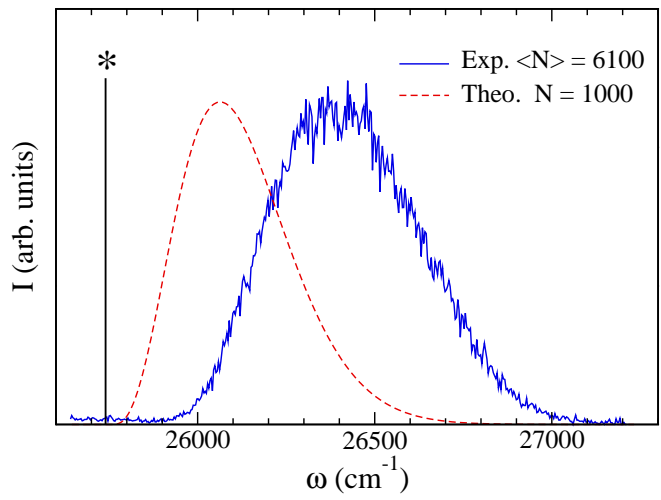}
\caption{\label{fig3} (Color online) Theoretical and experimental
excitation spectrum of the $4s \leftarrow 3s$ transition of Na on
$^4$He$_N$ droplets. The top-starred vertical line corresponds to the free
atom transition.\cite{NIST}}
\end{figure}

\begin{figure}[t]
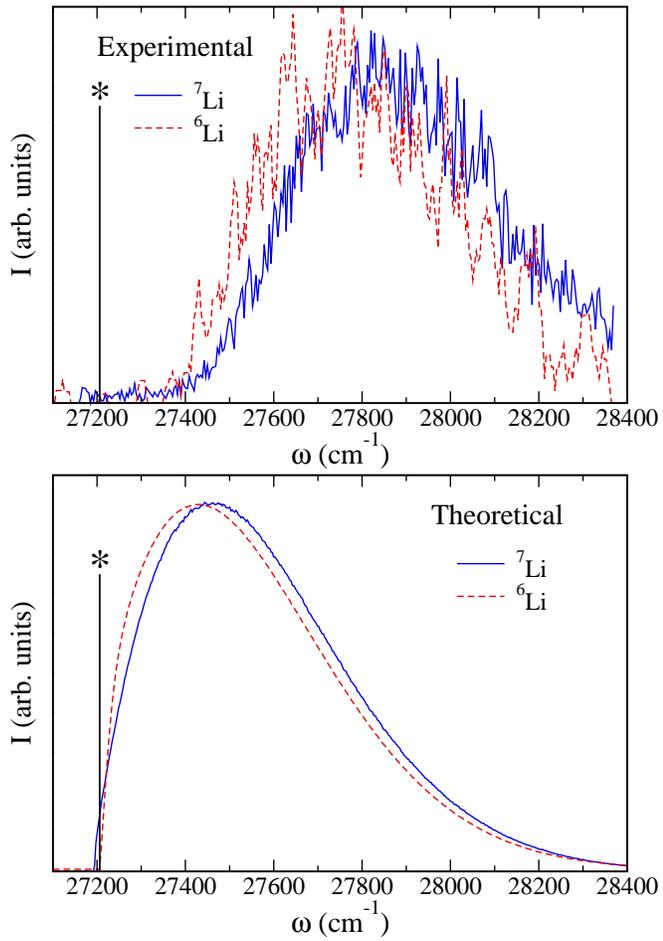

\includegraphics[width=\linewidth,clip=true]{speLiE.eps}
\includegraphics[width=\linewidth,clip=true]{speLiT.eps}
\caption{\label{fig4} (Color online) Top panel: Experimental
 excitation spectrum of the $3s \leftarrow 2s$ transition for Li
 attached to helium droplets consisting on average of 6100 atoms.
 Bottom panel: Corresponding theoretical spectrum for Li@$^4$He$_{1000}$.
 The top-starred vertical line corresponds to the free atom transition.\cite{NIST} }
\end{figure}

To investigate the desorption dynamics of sodium and lithium atoms
from the surface of helium nanodroplets, the alkali atoms have
been excited to their nominal $4s$ and $3s$ state, respectively.
The corresponding excitation spectra are shown in Figs. 3 and 4.
The spectra of the two lithium isotopes, $^6$Li and $^7$Li, have
been recorded individually by gating the detector at the
appropriate arrival time of these ions. It should be noted that
only bare sodium and lithium ions have been observed and that no
complexes with helium were detected, as was explicitly checked by
time-of-flight mass spectrometry. This implies that the excitation spectra reported in
Figs.~\ref{fig3} and  \ref{fig4}
 correspond to absorption spectra which allows for
a direct comparison between the experimental  and theoretical spectra.
The spectra are all characterized by a broad absorption
band that is blue-shifted with respect to the transition in the free
atom. The $4s \leftarrow 3s$ transition of sodium-doped droplets
has been discussed in detail before and the large blue-shift has
been attributed to the repulsive character of the
$4s$ effective potential,\cite{Log11} see also Fig.~\ref{fig1}.

The spectrum of the $4s \leftarrow 3s$ transition for the
Na@$^4$He$_{1000}$ system has been calculated taking explicitly
into account dimple fluctuations  using the atomic-like DFT
sampling technique, Eq.~(\ref{eqIdis6}). The width of the
experimental spectrum, 400 cm$^{-1}$, is well reproduced by the
calculations which yield a value of 370 cm$^{-1}$. In contrast,
the spectral shift, which is droplet size dependent and amounts to
570 cm$^{-1}$ for droplets consisting on average of 1700 helium
atoms, is somewhat underestimated by the
calculation that yields 350 cm$^{-1}$ for $N=1000$.

We would like to mention that although the absorption spectra of
Na\cite{Log11} and other impurities\cite{Bue07,Her08} calculated
by semiclassical methods yield at times quantitative agreement
with the experimental spectra, the DFT sampling method
employed here constitutes a consistent framework that is able to
reproduce the basic spectral features both in droplets and bulk
helium.\cite{Bue09,Her10,Her11,Mat11b} It is evident that since
the helium-helium correlations are described in a semiclassical
way, the method still needs some improvements. In spite of this
limitation, it is the only workable method for incorporating
density fluctuations within the DFT scheme, and it is for this
reason that we use it in this investigation.

\begin{figure}[p]
\includegraphics[width=\linewidth,clip=true]{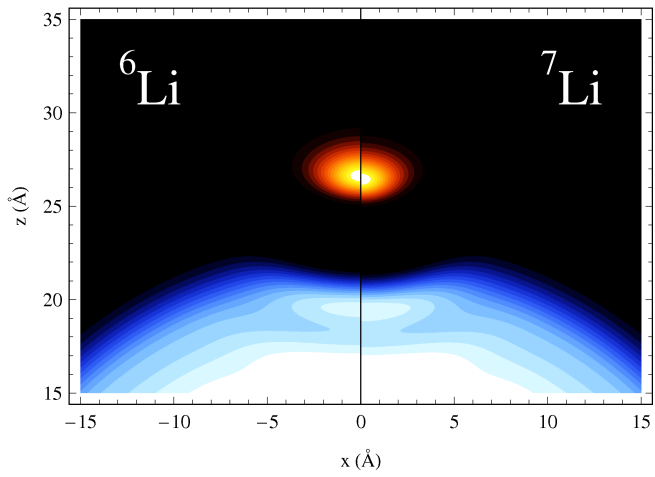}
\caption{\label{fig5} (Color online) Dimple structure of
$^6$Li@$^4$He$_{1000}$ (left) and $^7$Li@$^4$He$_{1000}$ (right)
droplets. The probability density distribution of the dopant is
also shown. }
\end{figure}

Inspection of the spectra for the two lithium isotopes shown in
Fig.~\ref{fig4} reveals that they are very similar but that the
spectrum of $^6$Li is slightly less blue-shifted than that of
$^7$Li. Solving the equivalent of Eqs. (\ref{eqH}) for both
isotopes indicates that this difference is related to the ground
state structure: as can be seen in Fig.~\ref{fig5} the lighter
$^6$Li isotope generates a less pronounced dimple structure and is
more delocalized than the heavier $^7$Li isotope. As a result, the
lighter isotope probes the excited state potential at larger
distances corresponding to lower energies. It is interesting to
note that similar zero point motion effects have been observed for
Li on $^3$He and $^4$He droplets.\cite{Her11} The
calculations accurately reproduce the width of the absorption line (540 cm$^{-1}$
in the calculations {\it vs.} 530 cm$^{-1}$ in the experiments) and
the isotopic shift (relative difference of 13\% in the
calculations and 15\% in the experiments). As discussed above,
the calculations somewhat underestimate the absolute blue-shift of the spectrum. We
would like to point out that since the helium-lithium interaction is
isotope-independent, all the differences found in the ground state
structure --and thus in the excitation spectra-- arise from the
kinetic energy term in Eqs. (\ref{eqH}) for Li. It is thus
a pure quantum effect (zero point motion) and consequently cannot be reproduced by the calculations
if the impurity is included as an external field.

\subsection{Photoelectron spectra}

\begin{figure}[p]
\includegraphics[width=\linewidth,clip=true]{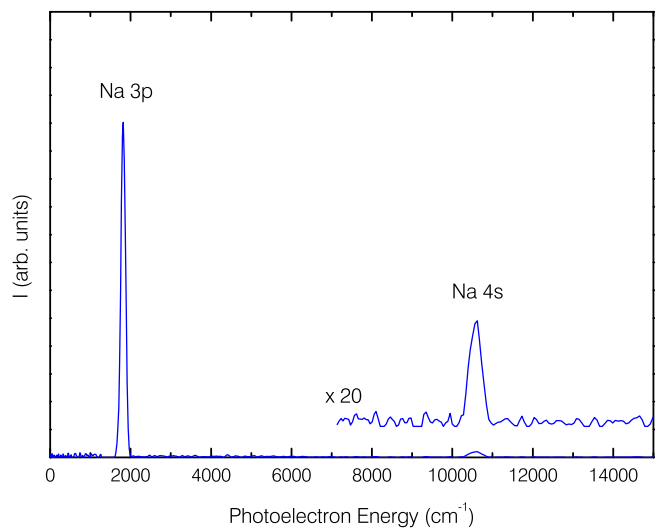}
\caption{\label{fig6} (Color online) Photoelectron spectrum
recorded following photoexcitation of sodium-doped helium droplets
at 26316 cm$^{-1}$. }
\end{figure}

The desorption efficiency and the  helium-induced relaxation
of excited alkali atoms have been investigated using photoelectron spectroscopy.
Fig.~\ref{fig6} shows the photoelectron spectrum obtained
following excitation of sodium-doped helium nanodroplets at a
frequency of 26316 cm$^{-1}$, corresponding to the maximum of the
absorption band. The spectrum is characterized by a strong peak at
low photoelectron kinetic energy and a much weaker peak at higher
energies. Based on the photon energy and the ionization
potential of sodium,\cite{NIST} the low energy peak can be readily
assigned to gas phase sodium atoms in the $3p$ state. The peak  at
high energy corresponds to free sodium in the $4s$ state. Since
the spectrum reveals no other peaks which could be assigned to
excited sodium atoms attached to the helium droplets,
\cite{Log11a} we conclude that all excited atoms desorb
from the droplets on the timescale of the laser pulse.

\begin{figure}[p]
\includegraphics[width=\linewidth,clip=true]{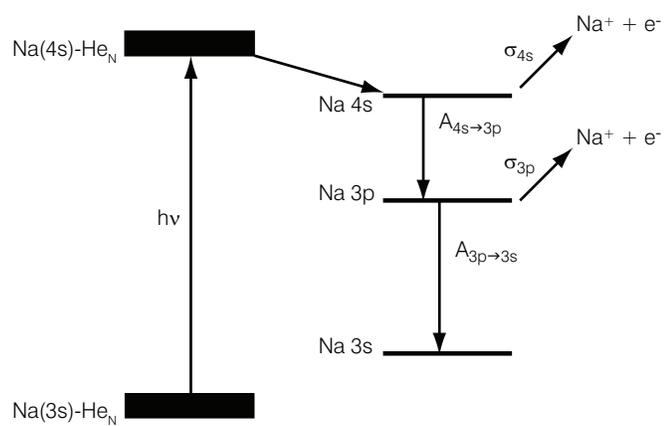}
\caption{\label{fig7} Energy level diagram displaying the
relation and excitation  processes taking place  after  excitation
of  sodium-doped helium droplets via the $4s \leftarrow 3s$
transition, see text for details. }
\end{figure}

The photoelectron spectrum gives the impression that the helium
induces a strong relaxation of the excited sodium, since the
majority of the sodium atoms are found to reside in the $3p$ state
and not in the initially excited $4s$ state. However, when
interpreting the photoelectron spectrum one has to take into
account that the $4s$ state has a short radiative lifetime \cite{NIST} and
that the ionization cross sections depend strongly on the excited
state of the sodium atom.\cite{Aym76,Aym78}
Unfortunately, it is not possible to address this issue by recording the
corresponding photoelectron spectrum of free sodium as the
$4s\leftarrow3s$ transition is dipole forbidden for one-photon
excitation in the free atom.
The photoelectron spectrum therefore has been simulated using a
rate equation approach. The model used is graphically depicted in
Fig.~\ref{fig7}. Following excitation of the surface bound sodium
atoms to the nominal $4s$ state they desorb from the helium
droplets. The free atoms decay by spontaneous emission to lower
lying states. This process is characterized by the Einstein
coefficients $A_{4s\rightarrow3p}$ and $A_{3p\rightarrow3s}$.
\cite{NIST} During the laser pulse the excited atoms are ionized
with an efficiency that is determined by the state specific
ionization cross sections $\sigma_{4s}$ and $\sigma_{3p}$
\cite{Aym76,Aym78} and the laser intensity. To model the
photoelectron spectra we now assume that no relaxation is induced
by the helium and that the atoms desorb instantaneously from the
droplets. This latter assumption is justified by the fact that the
radiative lifetime and the interaction time of the free atoms with
the light pulse is much longer than the desorption time of the
excited atoms, \emph{vide infra}. It should be noted that with
these assumptions the calculated photoelectron spectrum
corresponds to that of a free sodium atom resonantly ionized by a
$1+1$ photon excitation process via the $4s$ state. The set of
coupled equations describing the ionization of the free sodium
atoms has been solved numerically assuming that the 11 ns laser
pulse can de approximated by a Gaussian distribution. The
uncertainties of the various constants used, as well as  laser
power fluctuations have been included in the simulations to
determine the uncertainty of the relative intensities in the
calculated photoelectron spectrum.

The theoretically and experimentally determined intensity ratios are reported in
Table~\ref{tab1}. As can be seen from the table, the large intensity of the  $3p$
state is well reproduced by the calculations. This signifies that the
$3p$ state is mainly populated by spontaneous emission of $4s$
excited sodium atoms during the laser pulse. The small discrepancy
between the theoretical and experimental results might be
attributed to the large uncertainty associated with the
$\sigma_{4s}$ ionization cross section, which is close to its
minimum at the excitation frequency used. Alternatively,
it might indicate that some helium-induced relaxation takes places
during the desorption process. In view of the good agreement
between experiment and model calculations, we assume that all
sodium atoms leave the helium droplets in the initially excited
$4s$ state. Although no photoelectron spectra have been recorded
for lithium, we presume that they behave similar to sodium,
\emph{i.e.} all the excited lithium atoms desorb from the droplets
without undergoing helium-induced relaxation.

\subsection{Velocity and kinetic energy distributions}

\begin{figure}[p]
\includegraphics[width=\linewidth,clip=true]{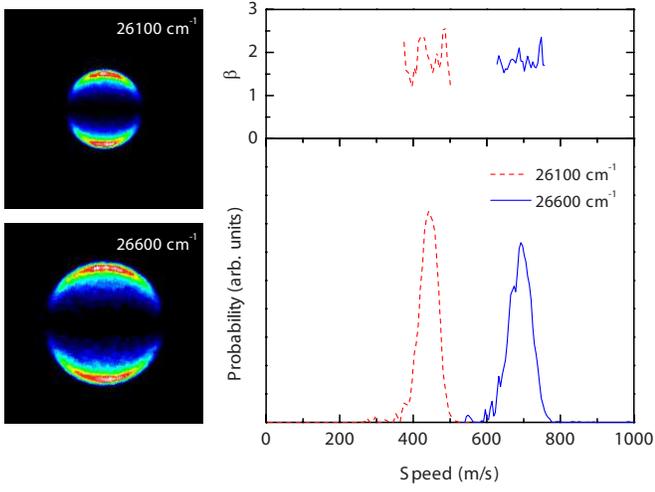}
\caption{\label{fig8} (Color online) Left: Ion images of sodium
atoms desorbed from the surface of helium droplets following
excitation at two frequencies within the $4s \leftarrow 3s$
resonance. The polarization of excitation laser is vertical with
respect to the images. Right: Speed distributions and anisotropy
parameters derived from the ion images.}
\end{figure}

To obtain insight into the desorption dynamics the velocity
distributions of the desorbed atoms have been determined. To this
end ion images have been recorded at several frequencies within
the absorption bands. Fig.~\ref{fig8} shows two ion images of
sodium that have been recorded following excitation at the low and high frequency
end of the $4s \leftarrow 3s$ excitation spectrum. Both images
are characterized by a strong anisotropic angular and radial
distribution, indicating that the desorbing atoms leave the
droplets with a well-defined velocity distribution. By performing
an inverse Abel transform to these images the speed distribution
and the angular anisotropy parameter $\beta$ have been determined.

The resulting speed distributions, which are also shown in Fig.~\ref{fig8}, are
found to depend strongly on the excitation frequency. While
excitation at 26100 cm$^{-1}$ yields sodium atoms with a most
probable speed of 440 m/s, excitation at 26600 cm$^{-1}$  yields
significant faster sodium atoms having speeds of 695 m/s. Whereas
the speed distributions depend on the excitation frequency, the
angular distributions do not show such dependence. The anisotropy
parameter $\beta$ is found to be independent of the velocity of
the desorbing sodium atoms and has a mean value of 1.81$\pm$0.06.
Similar values for the anisotropy parameter are found for desorbed lithium atoms, see
Table~\ref{tab2}. It should be noted that the speed distributions and the values for the anisotropy parameters
are found to be independent of helium droplet size. This can be attributed to the
local character of the interaction of the alkali atom with the helium droplet.

The values of the anisotropy parameter found in the experiments
are close to that expected for the parallel $(n+1)s\leftarrow ns$
transitions. As discussed before, the small deviation from this
value might be due to helium-induced configuration mixing.
Recently, Callegari and Ancilotto proposed a method to calculate
the interaction potentials of alkali atoms on helium nanodroplets
that explicitly takes into account configuration
mixing.\cite{Cal11} Using the expansion coefficients determined by
this method,\cite{Cal11a} evaluation of Eq. (\ref{betam}) yields a
value of $\beta=1.99$ for the anisotropy parameter. Evidently,
configuration mixing cannot account for the experimentally
observed anisotropy. A reduction of the anisotropy parameter also can
result if the rotational period of the helium droplets is
comparable to the time scale of desorption.\cite{Bus72a} Even
though not much is known about the rotation of helium droplets, it
is to be expected that the rotational period depends on the size
of the droplets. Since the anisotropy parameter is identical for
all three adatoms and does not depend on the velocity of the
desorbed atoms nor on the droplet size, it is highly unlikely that
droplet rotation is the cause for the reduced anisotropy
parameter. It is more likely that density fluctuations of the
helium in the proximity of the alkali atoms during the desorption
process lead to off-axis motion of the alkali atom. As we show in
Appendix A, the deformations induced by the helium fluctuations
can indeed result in a decrease of the anisotropy value.
An estimation of the order of magnitude of that decrease demands
an explicit calculation of the electronic configurations, which is
beyond the scope of this work.

\begin{figure}[p]
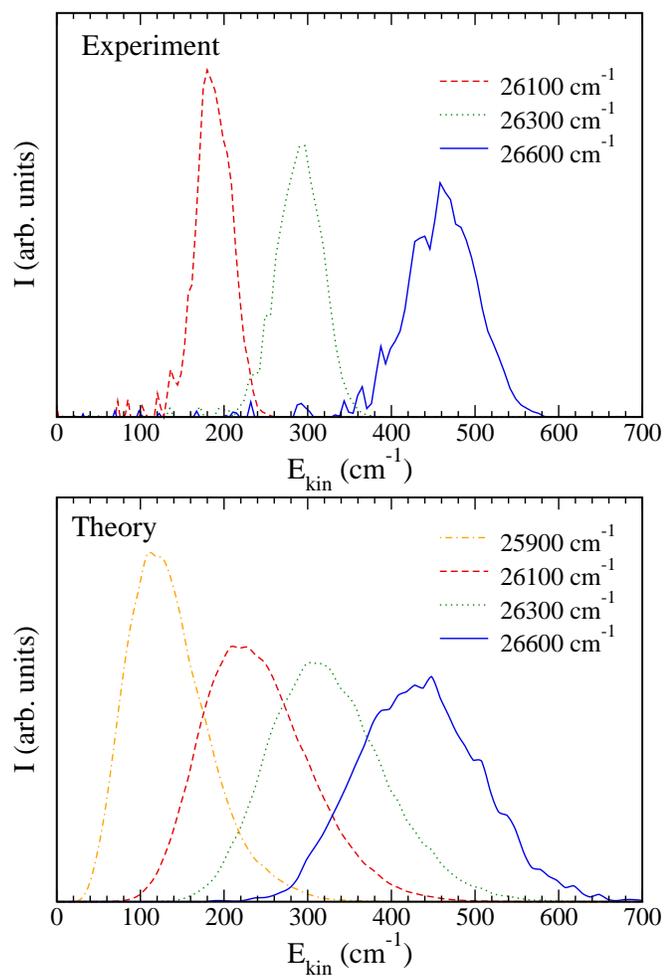

\includegraphics[width=\linewidth,clip=true]{kin_exp.eps}
\includegraphics[width=\linewidth,clip=true]{kin_the.eps}
\caption{\label{fig9} (Color online) Experimental (top panel) and
theoretical (bottom panel) normalized kinetic energy distributions
of desorbed Na atoms following excitation at different energies.}
\end{figure}

\begin{figure}[p]
\includegraphics[width=0.9\linewidth,clip=true]{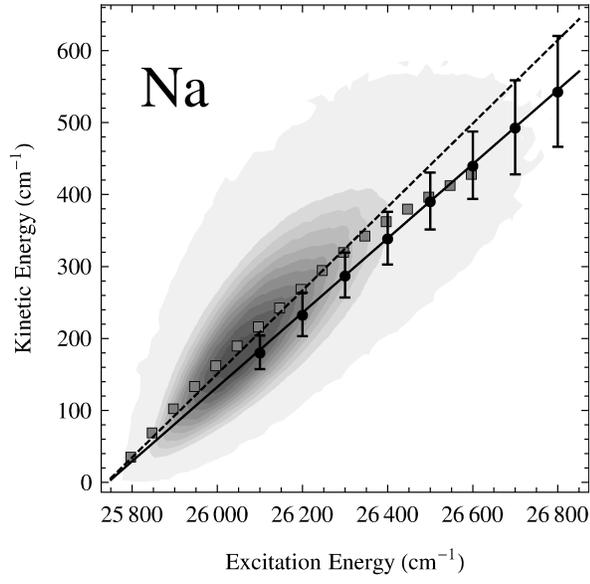}
\caption{\label{fig10} (Color online) $I(\omega,E_k)$ distribution
for Na atoms. Squares: theoretical mean kinetic energy. Dashed
line: linear fit to the theoretical data up to an excitation
energy of 26400 cm$^{-1}$.\cite{note}
Dots: experimental mean kinetic energy.
Bars: experimental standard deviation of the kinetic energy distributions. Solid line: linear fit to
the experimental data.}
\end{figure}

\begin{figure}[p]
\includegraphics[width=\linewidth,clip=true]{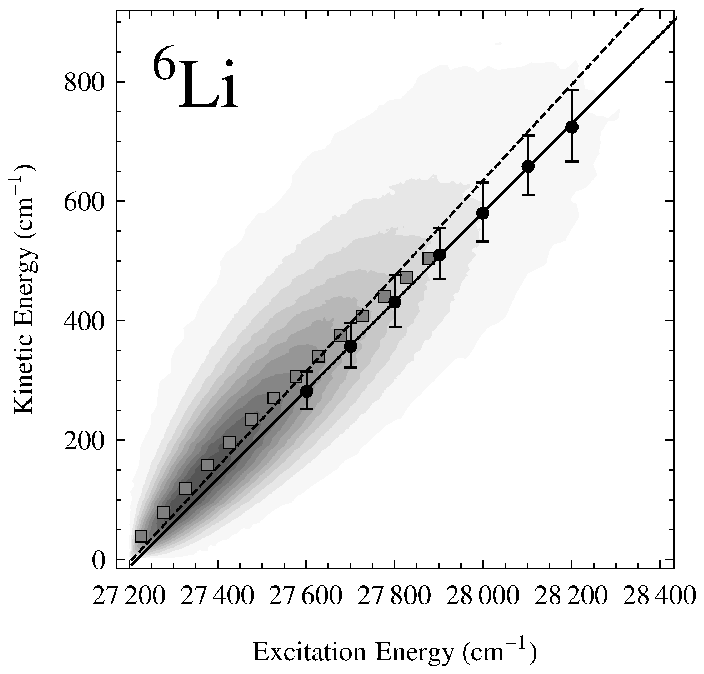}\\
\includegraphics[width=\linewidth,clip=true]{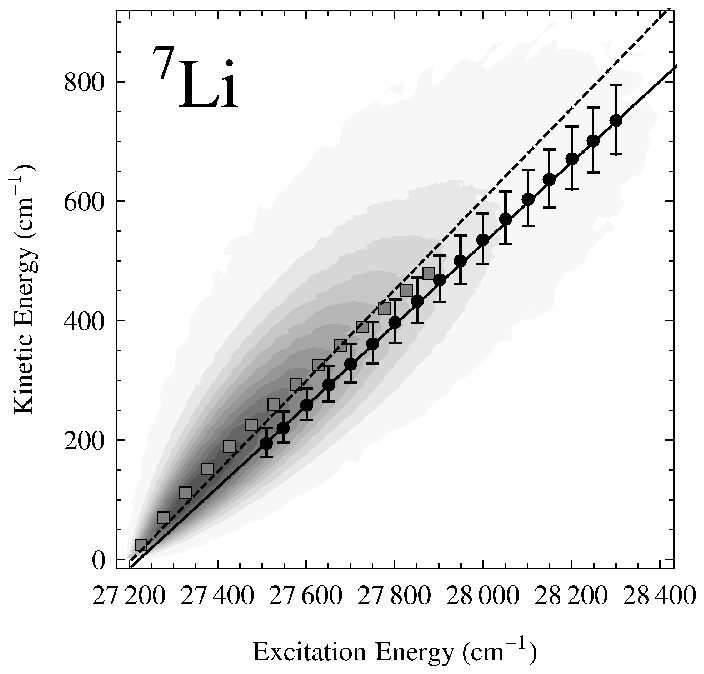}
\caption{\label{fig11} (Color online) Same as Fig. \ref{fig10} for
Li. In this case, all the points shown have been included in the
fit of the theoretical data.
 }
\end{figure}

In order to establish a relation between the kinetic energy of the
desorbing atoms and the excitation frequency, the speed
distributions have been transformed into kinetic energy distributions,
see Fig.~\ref{fig9}, and their average value and standard
deviation have been determined. Fig.~\ref{fig10} shows the average
kinetic energy of desorbed sodium atoms as function of excitation
frequency, while Fig.~\ref{fig11} shows
those for the two lithium isotopes.
For all three impurities the average kinetic energy
shows a linear dependence on the excitation frequency. The data
points therefore have been fitted to the following expression:
\begin{equation}\label{linear}
\langle E_{kin} \rangle =\eta (\hbar\omega-\hbar\omega_0)
\end{equation}
where $\hbar\omega_0$ corresponds to the excitation energy
yielding alkali atoms with zero kinetic energy, while $\eta$ is a
proportionality constant. The constants derived from the fits are
reported in Table~\ref{tab2}. The values of $\omega_0$  are very
close to the transition frequencies of the free atoms,
\emph{i.e.}, 25740 cm$^{-1}$ and 27206 cm$^{-1}$ for sodium and
lithium, respectively.\cite{NIST} The good agreement between these
values indicates that the proportionality constant $\eta$ can be
interpreted as the fraction of the available energy that is
converted into kinetic energy of the desorbing alkali atom. The
values of $\eta$ as determined in the experiment depend not only
on the type of alkali atom but also on its mass, see Table~\ref{tab2}. The lighter the
atom, the larger the value of $\eta$ and thus the larger the
fraction of available energy that is carried away by the departing
atom. The  standard deviation of the kinetic energy distribution,
indicated by the bars in Figs.~\ref{fig10} and \ref{fig11}, also
shows a linear dependence on excitation frequency and therefore
has been fitted to the same functional form as the average kinetic
energy. In contrast to the average kinetic energy, the
proportionality constant for the standard deviation,
$\eta_\Delta$, does not depend on the atom nor its mass, see
Table~\ref{tab2}.

More insight into the desorption process can be gained from the
calculations. As can be seen in Fig.~\ref{fig1}, excitation of Na
to the $4s$ state leads to a highly repulsive interaction between
the sodium atom and the droplet which triggers the desorption of
the impurity. Figure~\ref{fig12} shows the evolution of the
Na@$^4$He$_{1000}$ system after excitation of the sodium atom.
Inspection of this figure reveals that the desorption of the
sodium atom is accompanied by the creation of density waves in the
helium droplet. This indicates that a fraction of the energy
deposited in the system by the optical excitation is transferred
to the helium cluster. It can be seen from the figure that after
$\sim$1.5~ps the Na atom has left the droplet surface and the
helium starts filling the dimple. The calculations are stopped
after 5~ps, when the sodium atom has reached an asymptotic mean
velocity of $510$ m/s.
During the 5 ps time propagation, the position of the center-of-mass of the
droplet experiences a minute displacement of 0.1 \AA{}. Assuming a constant motion,
 the translational energy of the droplet is calculated to be only 0.7 cm$^{-1}$,
 indicating that most of the energy transferred to the helium cluster is
converted into internal energy.

In the calculations, the GR is evaluated according to Eq.
(\ref{eqIdis1}) using the initial position and final velocities of
the Na test particles as input. This expression gives direct
access to the final velocity distributions of the desorbed alkali
atoms. To obtain a correlation between the asymptotic kinetic
energy distributions and the excitation energy, the GR is
evaluated using Eq. (\ref{eqIdis2}). The results of the
calculations are compared with experimental kinetic energy
distributions in Fig.~\ref{fig9}. It can be seen that as the
excitation energy increases, the kinetic energy distributions
shift to higher energies and broaden. While these trends are
well reproduced by the calculations, the widths of the
distributions are somewhat overestimated. This can be attributed
to the semiclassical method used to simulate the density
fluctuations, as discussed before.

The full $I(\omega,E_k)$ distribution calculated for Na is shown in Fig.~\ref{fig10} by a false color representation.
The average kinetic energy of the desorbed sodium atoms derived from this distribution at selected excitation energies is also presented in this figure.
The calculations show a strong correlation between the mean kinetic energy and the
excitation energy, in agreement with experiments. A linear fit to
the theoretical data points yields a slope $0.58$ which compares
well to the experimental value of $0.52$. It is important to point
out that it is essential to take into account the density
fluctuations in order to achieve a good agreement with experiment.
If the density fluctuations are not included in the calculations a
slope of $0.76$ is obtained. This leads to the conclusion, that
the density fluctuations are responsible for the absorbtion of a large part of the energy deposited in
the system by the electronic excitation.\cite{Wad03} Calculations
for the smaller Na@$^4$He$_{500}$ complex yield results similar to
those for Na@$^4$He$_{1000}$. This size independence is in agreement
with experiment, and can be attributed to the local character of
the interaction of the alkali atom with the helium droplet.

The calculations reveal that lithium behaves very similar to sodium, although
there are some differences. Due to the lighter mass and
a less pronounced dimple, Li atoms desorb faster from the droplets
than Na atoms and reach the
asymptotic free regime already after $\sim$3~ps. Their asymptotic
mean velocities are significantly higher than for sodium and are
slightly different for the two Li isotopes, $^6$Li being faster
than $^7$Li. The $I(\omega,E_k)$ distributions  for Li are shown in
Fig.~\ref{fig11} together with the experimental data. Also in this
case the calculations display a similar correlation between the
mean kinetic energy of the desorbed atom and the excitation
energy. In agreement with experiment, see Table \ref{tab2},
slightly different slopes for the two isotopes are found, 0.80 for
$^6$Li and 0.76 for $^7$Li, confirming the mass-dependent nature
of the process.

As a final comment, we want to point out that while the
experimental mean kinetic energies vary essentially linearly with
excitation energy, the theoretical values show some non-linearity, especially at higher excitation energies.
We have found that the variation becomes more linear at these energies if we
do not include helium density fluctuations. This indicates that at least part of
the non-linearity arises from the way fluctuations are handled.

\subsection{Helium Density Waves}

\begin{figure}[p]
\includegraphics[width=\linewidth,clip=true]{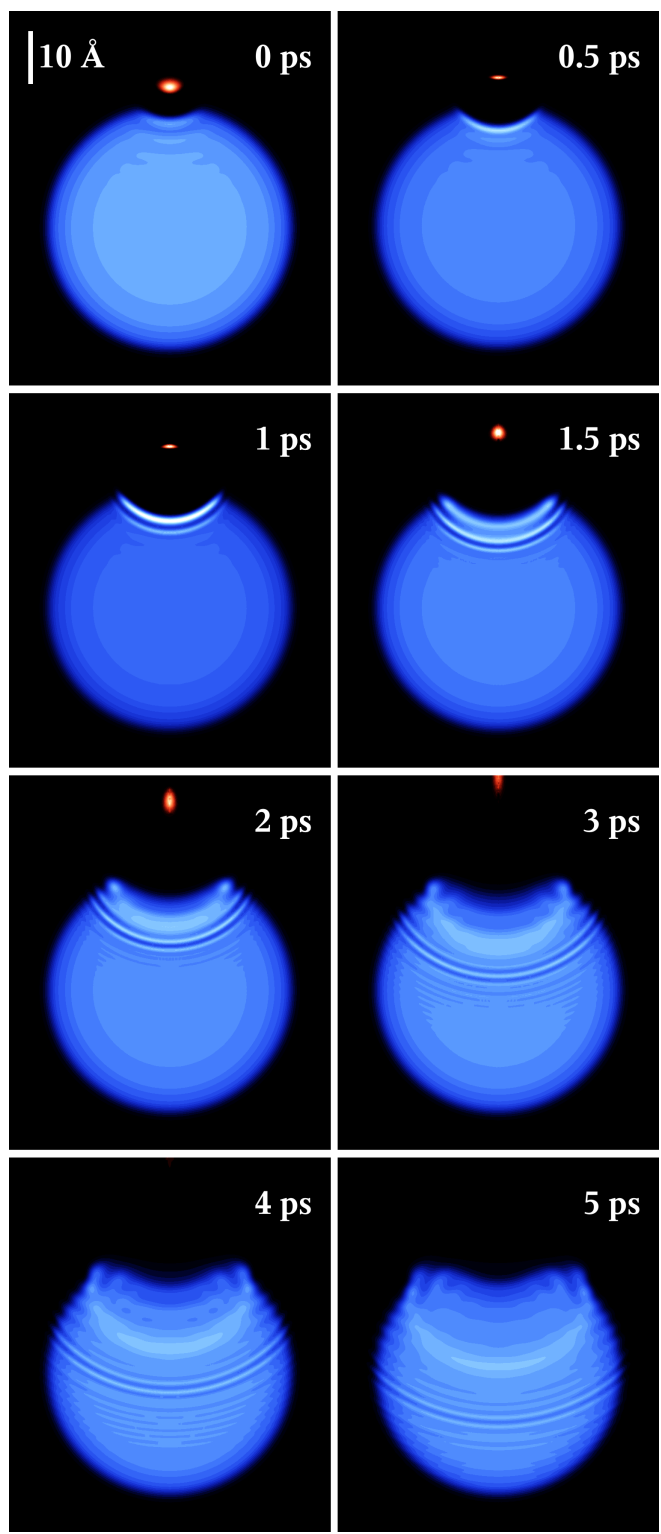}
\caption{\label{fig12} (Color online) Evolution of the
Na@$^4$He$_{1000}$ complex after excitation. }
\end{figure}

\begin{figure}[p]
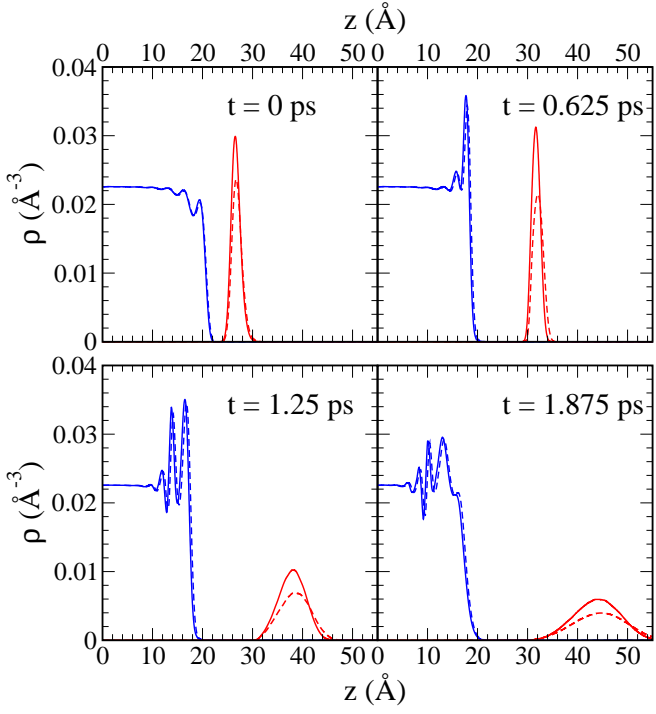

\includegraphics[width=\linewidth,clip=true]{evoLi1.eps}
\includegraphics[width=\linewidth,clip=true]{evoLi2.eps}
\caption{\label{fig13} (Color online) Helium density profiles and
Li probability density distributions (Gaussian-like profiles)
showing the dynamical desorption of Li isotopes along the
symmetry axis. Dashed lines, $^6$Li. Solid lines, $^7$Li. }
\end{figure}

Whereas the experiments provide only information on the desorbing
alkali atom, the calculations also provide insight into the
dynamics of the helium droplet upon excitation of
the adsorbed atom. As an example, in Fig.~\ref{fig13} the helium
density profile of a Li@$^4$He$_{1000}$ droplet along the symmetry
axis is represented as function of time. This figure, together with
Fig.~\ref{fig12} for Na, gives a pictorial yet quantitative
representation of the dynamics triggered by the excitation of the
alkali atom. In particular, they show the dramatic changes in the droplet density
caused by the excitation and subsequent desorption of the impurity.

\begin{figure}[p]
\includegraphics[width=\linewidth,clip=true]{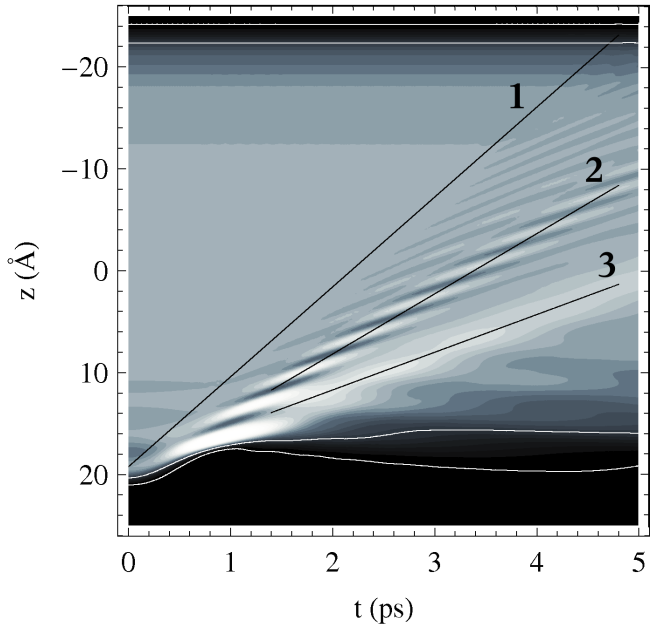}
\caption{\label{fig14} (Color online) Evolution of the helium
density profile of the Na@$^4$He$_{1000}$ system along the
symmetry axis. Three supersonic fronts are identified and
labeled by roman numbers. Equidensity lines corresponding to
0.5 and 0.1 times the helium saturation density, 0.0218 \AA$^{-3}$,
are shown in white.}
\end{figure}

To establish the nature of the helium density waves created by the
excitation, we concentrate on the simulations for Na, as they last
for longer times. Fig. \ref{fig14} shows the evolution of the
helium density profile of a Na-doped  $^4$He$_{1000}$ droplet
during the first 5 ps. Initially, the droplet extends along the
$z$ (symmetry) axis from about 21 to $-25$ \AA{}, and the Na atom
is located in a dimple at the surface. Excitation of the sodium to
the $4s$ state causes the dimple first to deepen due the highly
repulsive nature of the He-Na $4s$ interaction. The associated
compression of the helium last up to $\sim 1$ ps, as shown in the
figure. Following this compression, the helium surface bounces
back and the dimple starts being filled. The more distant part of
the droplet at $\sim -25$ \AA{} remains unperturbed and at rest,
indicating that energy deposited in the droplet leads almost
exclusively to the excitation of its internal degrees of freedom
and not to a translational motion of the droplet as a whole.

\begin{figure}[p]
\includegraphics[width=\linewidth,clip=true]{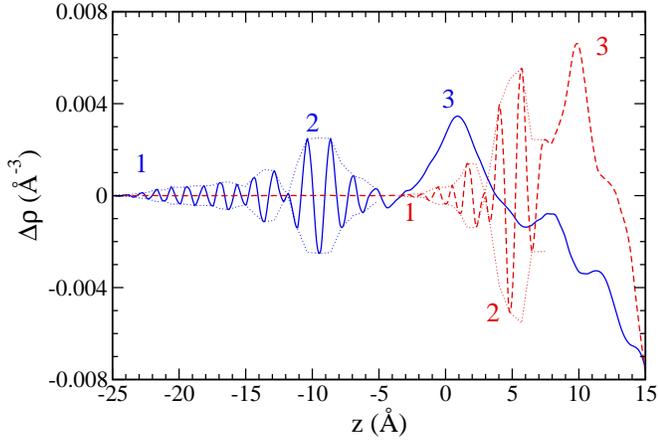}
\caption{\label{fig15} (Color online) Difference between the
density profile and the initial density along the symmetry axis,
$\Delta\rho \equiv \rho(t,z)-\rho(0,z)$, for $t=2.5$ ps (dashed
lines) and $t=5$ ps (solid lines). The envelope wave of the
modulated carrier waves is also shown for clarity (dotted lines).
The three identified fronts are labelled as in Fig. \ref{fig14}. }
\end{figure}

Figures \ref{fig12}-\ref{fig14} all reveal that excitation of the
alkali atom launches highly non-linear density waves into the
helium droplet. This becomes even clearer in Fig.~\ref{fig15}
which shows the density variations along the symmetry axis at
different times. The waves created by the excitation propagate in
the droplet at supersonic velocities.\cite{Wil67} In the case of Na, see
Figs. \ref{fig14} and \ref{fig15}, the first perturbation front,
labelled as 1, moves at $\sim$890 m/s and reaches the opposite
`edge' of the droplet in less than 5 ps. This perturbation
generates carrier waves with a phase velocity between 300 and 370
m/s, modulated by supersonic envelope fronts with growing
intensity. The ones with highest intensity, labelled as 2, have a
group velocity of $\sim$590 m/s. The origin of this modulation can
be traced back to the original structure of the droplet, being an
`echo' of the solvation shells around the impurity in its ground
state. Next, a high intensity wave appears travelling at $\sim$370
m/s (labelled as 3), which generates secondary waves propagating
backwards. A closer analysis of this wave has allowed us to
identify it with a solitary wave or soliton. Indeed, it can be
seen from Figs.~\ref{fig14} and \ref{fig15} that it always
corresponds to the maximum intensity of the travelling wave.

We have found similar density waves in the case of Li, for which we
recall that the evolution is stopped after 3 ps. This time is long
enough to establish that the first perturbation front moves at
about 750 m/s for both isotopes. This speed is lower than in
the case of sodium which might be attributed to the smaller amount of
energy that is deposited into the droplet.
Waves 2 and 3 develop as well, although they are not so clearly visible
and their velocities cannot be determined with confidence.

To our knowledge, the existence of these types of travelling waves
has not been disclosed before. They bear some similarities with
the waves produced in liquid helium by the
de-excitation of electron bubbles.\cite{Elo02,Mat11}
These waves have been identified with shock waves in Ref. \onlinecite{Elo02},
but with no clear justification for this statement.

\section{Further remarks}

The most pertinent result of this study is undoubtedly the fact
that the asymptotic kinetic energy of the desorbing alkali atoms
scales linearly with the excitation energy.
Both experiment and theory find
that the amount of energy carried away by the desorbing alkali atom
depends not only on the atom but also on its mass. In particular, the
larger the mass of the impurity, the larger the amount of energy
deposited in the helium droplet.

Even though the calculations reproduce the
experimental observations fairly well, they do not provide the physical
insight required to identify the
precise mechanism giving rise to the particular partitioning of
the available energy between the desorbing alkali atom and the
helium droplet. Rather than a limitation of the method used, this is a
signature of the actual complexity of the desorption process.

Nonetheless, one would like to have some insight into the physical
processes leading to the particular energy partitioning. In view of
the analogy, one might consider describing the desorption of an
excited alkali atom from the surface of a helium droplet as a
photodissociation reaction. Various models have been put forward
to describe the photodissociation of polyatomic molecules, each of
them focussing on specific aspects of the process.\cite{Mor97} In
case of a direct dissociation via excitation to a repulsive state,
the process is best described by the simple impulsive model
introduced by Busch and Wilson in their seminal work on the
photodynamics of NO$_2$.\cite{Bus72} The basic idea of the
impulsive model is that the force needed to break apart the
molecule is solely directed along the axis of the dissociating
bond. Under the assumption that the bond breaks promptly, the
available energy is initially partitioned, by conservation of
linear momentum, between the kinetic energies of the two atoms
forming the bond. The kinetic energy of the atoms is subsequently
partitioned between the translation, rotational and vibrational
degrees of freedom of each fragment. This impulsive model might be
well suited to describe the desorption of an alkali atom from the
surface of a helium droplet given that excitation of the alkali
atom leads to a highly repulsive interaction between the excited
atom and the helium droplet, see Fig.~\ref{fig1}, and a fast
desorption of the alkali atom. The main
difficulty in applying the impulsive model to the helium droplet
system lies in that the alkali atom is not bound to a
single helium atom but interacts with all the atoms making up the
droplet. However, due to its short range character, the repulsive
interaction is by far the strongest with the nearby helium atoms
located at the surface of the dimple.\cite{Pas83} This is also
born out by the calculations which reveal that predominantly these
helium atoms are displaced immediately after excitation of the
alkali atom, see Fig.~\ref{fig12}. One might therefore consider
reducing the problem to a pseudo-polyatomic model in which the
alkali atom having a mass of $m_{\mathrm{Ak}}$ is considered to be
bound to the helium droplet via a single helium moiety with an
effective mass of $m_{\mathrm{eff}}$ that is to be determined.
Assuming that the impulsive model can be applied to this strongly
simplified description of the system, the kinetic energy of the
desorbed alkali atom, $E_{kin}({\rm Ak})$ is related to the available
energy $\Delta\hbar\omega$ according to:
\begin{equation}
E_{\mathrm{kin}}(\mathrm{Ak})=
\frac{m_{\mathrm{eff}}}{m_{\mathrm{eff}}+m_{\mathrm{Ak}}}\Delta\hbar\omega
\; .
\label{slope}
\end{equation}
The model thus yields a linear dependence of the kinetic energy on
excitation frequency, as is observed in the present study. The
slope is directly related to the effective mass of the helium
moiety with which the excited alkali atom interacts. Since the
alkali atom interacts with several helium atoms located at the
surface of the dimple, we expect $m_{\mathrm{eff}}$ to be larger
than the mass of a single helium atom. Although $m_{\mathrm{eff}}$
cannot be calculated \emph{a priori}, it can be determined by
fitting the experimental slope of the kinetic energy on excitation
frequency, see Table \ref{tab2}. One finds an effective mass of
$\sim$16 and $\sim$24 amu for Li and Na, respectively. This would
signify that the $3s$ electron of Li interacts on average with 4
helium atoms, while the $4s$ electron of sodium interacts with 6
helium atoms. The difference in the number of interacting helium
atoms is thought to reflect the difference in electron orbit
radius and dimple structure.

We would like to stress here that in spite of the fact that the
impulsive model offers an explanation for some of the experimental
and theoretical observations, it is only an approximative
description that lacks any predictive power. The effective mass
required for this model can only be determined \emph{a posteriori}
from the experimental or theoretical data. Due to the
approximative nature of the model this effective mass should be
interpreted with care. This is exemplified by the results for the
two lithium isotopes, where according to the impulsive model
$^6$Li interacts with more helium atoms than $^7$Li, see
Table~\ref{tab2}. This contradicts the conclusion based on the
difference in the excitation spectrum of the two lithium isotopes and the calculations of
the ground state structure,
which indicate that the lighter $^6$Li atom
is located further away from the droplet surface and consequently
interacts less with the helium than the $^7$Li atom.

As a final remark, the weak dependence of the standard deviation
of the energy distributions on the alkali atom or its mass points
to the helium as the source of the broadening. As seen in the
calculations, helium density fluctuations cause some dispersion in
the observables. Since these fluctuations are to a large extent
independent of the impurity attached to the droplet, it is
expected that so is the dispersion of the observables.

\section{Summary}

We have carried out a combined experimental and theoretical
investigation of the desorption of  Na and Li alkali atoms from
the surface of helium droplets following excitation via the
$(n+1)s \leftarrow ns$ transition. These systems are well suited
to gain insight into the dynamics of this complex phenomena, since
neither exciplex formation nor helium-induced relaxation of the
impurity obscure the analysis of the experimental findings or its
theoretical interpretation.\cite{Log11} Additionally, the use of
Li allows to address the mass effect in the desorption process by
making a direct comparison between the results for $^6$Li and
$^7$Li.

The analysis of the experimental results has been carried out
within a full dynamical, three dimensional approach that combines
a time-dependent DFT description of the droplet with a Bohmian
description of the impurity. To the best of our knowledge, this
is the first time that such a theoretical framework has been developed and
applied to the desorption of impurities from helium droplets.

The experiments reveal that the $(n+1)s \leftarrow ns$ transitions
of Li and Na atoms located on helium droplets are significantly
blue-shifted with respect to the corresponding gas phase
transitions. They furthermore disclose that excitation of the
alkali atom leads in all cases to its desorption from the helium
droplet and that the average kinetic energy of the desorbed atom
depends linearly on the excitation energy. These observations are
all reproduced by the calculations, which allows us to have
confidence in theoretical observations that cannot be experimentally
verified. More specifically, the calculations indicate that
the energy deposited in the system by the excitation of the alkali
atom leads to the creation of highly non-linear helium density
waves that propagate through the helium droplet at supersonic
velocities. One of such waves could be identified as a soliton.
Based on the good agreement between experiment and theory it
becomes also possible to identify the main physical ingredients
necessary for a quantitative description of the desorption
process, namely: i) the zero-point motion of the impurities, ii) a
full dynamical description of both the helium droplet and the
impurity, and iii) the inclusion of helium density fluctuations.
These concepts have been implemented in a numerical approach that
is very robust and can be applied to the description of other
dynamical processes involving atomic impurities propagating in
helium droplets.\cite{Bon07,Bon08,Log07,Bra07,Bra04} Addressing
exciplex formation, as observed in several
systems,\cite{Per96,Reh00,Tak03,Tak04,Dro04,Eno04,Bra07a,Log11}
however, still remains a challenge from a computational point of view that will
likely require a test particle description of both the impurity
and the helium droplet.

\section*{Acknowledgments}

We would like to thank David Mateo for useful discussions. This
work has been performed under Grants No. FIS2008-00421/FIS from
DGI, Spain (FEDER), 2009SGR1289 from Generalitat de Catalunya, and
200020-119789 from the Swiss National Science Foundation. AH has
been supported by the ME (Spain) FPI program, Grant no.
BES-2009-027139.

\begin{widetext}
\appendix
\section{}

We discuss in this Appendix the probability of an optical transition
between mixed electronic orbitals. Defining the
initial and final states in cylindrical symmetry as
\begin{eqnarray}
\displaystyle\varphi_i(\theta)&=\displaystyle\sum_{\ell=0}^\infty A_{\ell}Y_{\ell0}(\theta)\nonumber\\
\displaystyle\varphi_f(\theta)&=\displaystyle\sum_{\ell=0}^\infty B_{\ell}Y_{\ell0}(\theta)\nonumber,
\end{eqnarray}
where $Y_{\ell m}$ are the spherical harmonics of order
($\ell,m$), the angular dependence of the dipolar transition probability
is written as\cite{zare}
\begin{equation}
P_{i\rightarrow f}(\theta)=\sum_{\ell\ell'}|B_\ell|^2|A_{\ell'}|^2|\langle\ell'010|\ell0\rangle|^2
|Y_{\ell0}(\theta)|^2
\end{equation}
where $\langle\ell'010|\ell0\rangle$ is a Clebsch-Gordan
coefficient. Due to the properties of these coefficients we can write
\begin{equation}
P_{i\rightarrow f}(\theta)=
\sum_{\ell}|B_\ell|^2\frac{|A_{\ell-1}|^2\ell(3+2\ell)+|A_\ell+1|^2(2\ell^2
+\ell-1)}{4\ell(1+\ell)-3}|Y_{\ell0}(\theta)|^2\equiv\sum_{j=0}^\infty D_j P_{2j}(\cos\theta)
\end{equation}
where $P_j(x)$ are the Legendre polynomials of order $j$. If the
involved states are limited to $\ell\leq1$
\begin{eqnarray}
\displaystyle\varphi_i(\theta)&\displaystyle=Y_{00}+A_1Y_{10}(\theta)\nonumber\\
\displaystyle\varphi_f(\theta)&\displaystyle=Y_{00}+B_1Y_{10}(\theta)\nonumber,
\end{eqnarray}
we obtain for the transition probability
\begin{equation}\label{eqAb}
P_{i\rightarrow
f}(\theta)=\frac{|A_1|^2}{12\pi}+\frac{3|B_1|^2}{4\pi}\cos^2(\theta)
\propto 1+\frac{2}{1+|A_1|^2/(3|B_1|^2)}P_2(\cos\theta) \equiv
1+\beta P_2(\cos\theta) \label{betam}
\end{equation}
It can be seen that the anisotropy can only take positive values.
When $|B_1|^2>|A_1|^2$, the anisotropy is constrained to
$1.5<\beta<2$, and  if the initial state
is spherical ($A_1=0$) one finds the limiting value $\beta=2$.

When non-axially symmetric deformations are included ($m\neq0$) the
transition probability for $\ell\leq1$ reads

\begin{equation}
P_{i\rightarrow f}(\theta)=\sum_{\ell\ell'\leq1,mm'}|B_{\ell
m}|^2|A_{\ell'm'}|^2|\langle\ell'm'10|\ell m\rangle|^2
|Y_{\ell0}(\theta)|^2\propto 1+\beta_0K P_2(\cos\theta)
\end{equation}
with $\beta_0=2/(1+|A_{10}|^2/(3|B_{10}|^2))$ [same structure as
in Eq. (\ref{eqAb})] and

\begin{equation}
K = \frac{ (|A_{10}|^2 + 3 |B_{10}|^2) (4 |B_{10}|^2 -
|A_{11}|^2|B_{11}|^2 - |A_{1-1}|^2|B_{1-1}|^2) }{ 2 |B_{10}|^2
[2(|A_{10}|^2 + 3 |B_{10}|^2) +    3
(|A_{11}|^2|B_{11}|^2+|A_{1-1}|^2|B_{1-1}|^2)] }.
\end{equation}
Some algebra shows that $K<1$, since
$(|A_{10}|^2 + 9 |B_{10}|^2)
(|A_{11}|^2|B_{11}|^2+|A_{1-1}|^2|B_{1-1}|^2) > 0$.
Consequently, these deformations will
always decrease the value of the anisotropy parameter.

\section{}

We show in this appendix how we have represented the delta
functions involved in Eq.~(\ref{eqIi}) to compute the GR. Using
the rectangular function $\text{rect}(x)\equiv
\vartheta(x+1/2)-\vartheta(x-1/2)$, where $\vartheta(x)$ is the
Step function, we write
\begin{equation}
\label{eqIdis1}
w_{i\rightarrow f} \propto
\left[1+\beta P_2(\cos\theta)\right]
\frac{1}{M}\sum_{i=1}^M\text{rect}\!\left[\frac{\hbar\mathbf{k}-m_{\mathrm{Na}}
\dot{\mathbf{R}}_i(t_\infty)}{\Delta k}\right]\frac{1}{\Delta k}
\text{rect}\!\left[\frac{\hbar\omega-
\left[V_{\mathrm{Na}}^{4s}(0,\mathbf{R}_i(0))-\hbar\omega_{gs}\right]}
{\Delta\omega}\right]\frac{1}{\Delta\omega}  \; ,
\end{equation}
where we have used for the number of test particles $M=200000$ and
the intervals take small values as $\Delta k/m_{\mathrm{Na}}\sim5$
m/s and $\Delta\omega\sim3$ K. The velocity distributions are
simulated by choosing a value of the excitation energy $\omega$
and evaluating Eq.~(\ref{eqIdis1}), while the final kinetic energy
{\it vs.} excitation energy distribution is obtained  after the
change of variable $E_k=\hbar^2k^2/2m_{\mathrm{Na}}$ as
\begin{equation}
\label{eqIdis2}
\begin{array}{rl}
\displaystyle I(\omega,E_k)&=\displaystyle
\sqrt{\frac{m_{\mathrm{Na}}}{2E_k \hbar^2}}\,
I\left(\omega, k=\sqrt{\frac{2m_{\mathrm{Na}}E_k}{\hbar^2}}\right)\\
&\displaystyle =\frac{1}{M}\sum_{i=1}^M\text{rect}\!\left[\frac{E_k-m_{\mathrm{Na}}
\dot{\mathbf{R}}_i^2(t_\infty)/2}{\Delta E_k}\right]\frac{1}{\Delta E_k}
\text{rect}\!\left[\frac{\hbar\omega-
\left[V_{\mathrm{Na}}^{4s}(0,\mathbf{R}_i(0))-\hbar\omega_{gs}\right]}
{\Delta\omega}\right]\frac{1}{\Delta \omega}  \; .
\end{array}
\end{equation}
The excitation spectrum is computed as
\begin{equation}\label{eqIdis3}
I(\omega)=\frac{1}{M}\sum_{i=1}^M
\text{rect}\!\left[\frac{\hbar\omega-
\left[V_{\mathrm{Na}}^{4s}(0,\mathbf{R}_i(0))-\hbar\omega_{gs}\right]}
{\Delta\omega}\right]\frac{1}{\Delta\omega}.
\end{equation}
If density fluctuations are included using the stochastic  method
described in detail in Refs. \onlinecite{Her10,Mat11b}, the total distribution is
generated by the contribution of $N_c$ configurations generated by
sorting $N$ random positions in the $j$-th configuration
$\{\mathbf{r}^j_n\}_{n=1}^{N}$ for the hard spheres that represent
the $N$ helium atoms, using the helium density divided by $N$ as
probability density distribution together with a hard-sphere
repulsion between He atoms. The diameter of the sphere is of the
order of the length $h$ used to screen the Lennard-Jones potential
and to compute the coarse-grained density.\cite{Dal95} This
diameter is defined as\cite{Her10,Mat11b}
\begin{equation}
d^j_n=h\left(\frac{\rho_0}{\bar{\rho}(\mathbf{r}^j_n)}\right)^{1/3} \; ,
\end{equation}
where $\rho_0$ is the liquid saturation density value, and
$\bar{\rho}$ is the coarse-grained density obtained by averaging
the atomic density within a sphere of radius $h$.\cite{Dal95} The
distributions are then obtained as
\begin{equation}
\label{eqIdis4}
w_{i\rightarrow f} \propto\frac{1}{N_c}\sum^{N_c}_{j=1}\sum^N_{n=1}
\left[1+\beta P_2(\cos\theta)\right]
\frac{1}{M}\sum_{i=1}^M\text{rect}\!\left[\frac{\hbar\mathbf{k}-m_{\mathrm{Na}}
\dot{\mathbf{R}}_i(t_\infty)}{\Delta k}\right]\frac{1}{\Delta k}
\text{rect}\!\left[\frac{\hbar\omega-\left[V^{4s}_X(\left|\mathbf{r}^j_n
-\mathbf{R}_i(0)\right|)-\hbar\omega_{gs}\right]}
{\Delta\omega}\right]\frac{1}{\Delta\omega},
\end{equation}
\begin{equation}
\label{eqIdis5}
I(\omega,E_k)=\frac{1}{N_c}\sum^{N_c}_{j=1}\sum^N_{n=1}\frac{1}{M}
\sum_{i=1}^M\text{rect}\!\left[\frac{E_k-m_{\mathrm{Na}}
\dot{\mathbf{R}}_i^2(t_\infty)/2}{\Delta E_k}\right]\frac{1}{\Delta E_k}
\text{rect}\!\left[\frac{\hbar\omega-\left[V^{4s}_X(\left|\mathbf{r}^j_n
-\mathbf{R}_i(0)\right|)-\hbar\omega_{gs}\right]}
{\Delta\omega}\right]\frac{1}{\Delta \omega}  \; ,
\end{equation}
and
\begin{equation}\label{eqIdis6}
I(\omega)=\frac{1}{N_c}\sum^{N_c}_{j=1}\sum^N_{n=1}\frac{1}{M}\sum_{i=1}^M
\text{rect}\!\left[\frac{\hbar\omega-
\left[V^{4s}_X(\left|\mathbf{r}^j_n-\mathbf{R}_i(0)\right|)-\hbar\omega_{gs}\right]}
{\Delta\omega}\right]\frac{1}{\Delta\omega}.
\end{equation}
Note that with a number of configurations $N_c=10000$, the
histograms are computed  using a total of $N_c\times M=2~10^9$
contributions, large enough to reduce the statistical noise.

\end{widetext}

%
%
\clearpage

\clearpage

\clearpage

\clearpage

\clearpage

\clearpage

\clearpage

\clearpage

\clearpage

\clearpage

\clearpage

\clearpage

\clearpage

\clearpage

\clearpage

\clearpage
%
%

\begin{table}[p]
\begin{tabular}{c c  c c c}
\hline \hline
State & \hspace{1.0 cm} &  Experiment & \hspace{1.0 cm} &   Model  \\
\hline
3p   & &  0.96(1)  &   &   0.89(4)  \\
4s   & &  0.04(1)  &   &   0.11(4)  \\
\hline \hline
\end{tabular}
\caption{\label{tab1} Experimental and calculated relative intensities of the
photoelectron peaks following excitation of sodium-doped helium droplets via
the $4s \leftarrow 3s$ transition. }
\end{table}

\clearpage

\begin{table}[p]
\begin{tabular}{c c c c c c c c c c c c c c c}
\hline \hline
& &  \multicolumn{9}{c}{Experiment}  & & \multicolumn{3}{c}{Theory}
 \\
 \cline{3-11} \cline{13-15} \\ [-3.0ex]
Atom   &  \hspace{0.5 cm}    & $\omega_0$ [cm$^{-1}$] & \hspace{0.5 cm}  & $\eta$  & \hspace{0.5 cm}
& $\eta_\Delta $  & \hspace{0.5 cm} & $\beta$ & \hspace{0.5 cm} & m$_{\mathrm{eff}}$ [amu] &\hspace{0.5 cm}
& $\omega_0$ [cm$^{-1}$]\cite{NIST} &\hspace{0.5 cm} & $\eta$  \\ [0.5ex]
\hline
$^6$Li      & & 27218(6)   & & 0.743(6)  &  &  0.042(6) & & 1.73(6) & & 17.4 & & 27206  & & 0.802(8)  \\
$^7$Li      & & 27222(3)   & & 0.687(3)  &  &  0.040(2) & & 1.79(3) & & 15.4 & & 27206  & & 0.756(8)  \\
$^{23}$Na   & & 25743(4)   & & 0.516(4)  &  &  0.038(3) & & 1.81(6) & & 24.5 & & 25740  & & 0.583(9)  \\ [0.5ex] \hline
\hline
\end{tabular}
\caption{\label{tab2} Characteristics of the experimental and theoretical kinetic energy distributions
of the desorbed alkali atoms, see text for details. }
\end{table}


\begin{thebibliography}{99}
\bibitem{note}
Above an excitation energy of 26400 cm$^{-1}$ the data points clearly deviate from the linear dependence observed at lower energies and therefore have not been included in the fit.
\end{thebibliography}
\end{document}